\documentclass[fleqn,12pt,twoside]{article}
\usepackage{espcrc1}

\usepackage{graphicx}
\usepackage[figuresright]{rotating}

\newcommand \beq{\begin{eqnarray}}
\newcommand \eeq{\end{eqnarray}}  
\newcommand \li{\par\noindent}

\newcommand{\AmS}{{\protect\the\textfont2
  A\kern-.1667em\lower.5ex\hbox{M}\kern-.125emS}}

\title{Neutrino-nucleon scattering rate in the relativistic random 
phase approximation}

\author{L. Mornas\address[U-ovi]{Departamento de F{\'\i}sica, 
        Universidad de Oviedo, E33007 Oviedo, Spain}%
        \thanks{e-mail: lysiane{\char64}pinon.ccu.uniovi.es 
         \newline
         Work supported in part by the Spanish Grant n$^o$ 
         MCT-00-BFM-0357}}
       
\begin{document}

\maketitle

\begin{abstract}
The in-medium modification to the neutrino-nucleon scattering rate 
is calculated in the relativistic random phase approximation
in the framework of a hadronic meson exchange model, in view of 
applications to neutrino transport in supernovae and protoneutron stars.
\end{abstract}

\section{Introduction}

Neutrino transport is an important ingredient of the
numerical simulation of supernova collapse and of the cooling of the 
resulting protoneutron star.  
In the conditions of high density and temperature prevailing in this 
situation, the neutrino opacities will be appreciably modified by 
medium effects. This has been confirmed by the calculations of 
several groups (see {\it e.g.} \cite{RPLP99,YT00,MP02} and references
therein).
\li
We perform here a calculation of the effect of correlations on this
parameter in the random phase approximation. This contribution is an 
extension of the results presented in \cite{MP02} to the more commonly 
used parametrizations of the nuclear interactions NL3 and TM1
\cite{NL3,TM1}. The reader is referred to \cite{MP02} for detailed 
analytical expressions as well as for a more complete list of 
bibliographical references.
\li
In the dense and hot nuclear matter present in the protoneutron star, 
the mean free path of the neutrino is determined mostly by its interactions 
with the baryons. Here we will consider only the nucleons. The scattering 
process occurs {\it via} the neutral current, while the emission and 
absorption processes involve the charged current. In both cases, the 
differential cross section is given by
\beq
{d \sigma \over d E_\nu d \Omega} = {G_F^2 \over 64 \pi^3} 
{E_{\nu '} \over E_\nu} {\cal I}m (S^{\mu\nu} L_{\mu\nu})
\eeq 
It is proportional to the imaginary part of the lepton current 
$L^{\mu\nu}$ contracted with the structure function of the nucleons
$S^{\mu\nu}$. When the nucleons are free, the structure function
is proportional, with a thermodynamical factor, to the mean field value 
of the polarization
\beq 
\Pi_{WW}^{\mu\nu\, \mbox{\rm{\small MF}}}(q) = \int d^4p \ {\rm Tr} 
\left[ \Gamma_W^\mu G(p) \Gamma_W^\nu G(p+q) \right]
\eeq
where $G$ is the propagator of the nucleon and $\Gamma_W$ is the vertex 
for the weak interaction. 
\li
The RPA correlations are introduced by replacing 
the mean field polarization by the solution of the Dyson equation 
\beq
\noalign{\vskip -0.05cm}
\Pi_{WW}^{\mu\nu\, \mbox{\rm \small{RPA}}} =  \Pi_{WW}^{\mu\nu\, 
\mbox{\rm \small{MF}}} + \Pi_{WS}^{\mu\alpha\, \mbox{\rm \small{MF}}}\ 
D_{SS\, \alpha\beta}^{(0)}  \Pi_{SW}^{\beta\nu\, \mbox{\rm \small{RPA}}}
\eeq
\vskip -0.05cm \li
with $D_{SS}^{(0)}$ representing the nuclear interaction in free 
space\footnote{The indices $W$ or $S$ stand for a vertex with a weak 
or strong coupling respectively.}
As can be easily seen on the diagramatical representation, this 
corresponds to an infinite series of loop diagrams, which may also 
be resummed by defining a screened nuclear interaction
$D_{SS}^{\rm{\small RPA}}$
\beq
\noalign{\vskip -0.05cm}
\Pi_{WW}^{\mu\nu\, \mbox{\rm{\small RPA}}} =  \Pi_{WW}^{\mu\nu\, 
\mbox{\rm{\small MF}}} + \Pi_{WS}^{\mu\alpha\, \mbox{\rm{\small MF}}}\ 
D_{SS\, \alpha\beta}^{\mbox{\rm{\small RPA}}}\  \Pi_{SW}^{\beta\nu\, 
\mbox{\rm{\small MF}}}
\eeq

\section{Polarization insertions}
We will describe the nuclear interaction by a relativistic field model 
with meson exchange. Non-linear meson-meson interactions are introduced 
in order to obtain a better description of the properties of 
nuclear matter at saturation. We consider here two widely used 
parametrizations of the nuclear interaction, NL3 and TM1 \cite{NL3,TM1}, 
which are fitted to the properties of nuclei and bulk nuclear matter. 
The interaction piece of the corresponding Lagrangian reads
\beq
\noalign{\vskip -0.05cm}
{\cal L}_{\rm int} =
\overline\psi \left(-g_\sigma \sigma +g_\omega \gamma^\mu \omega_\mu 
+ g_\rho \gamma^\mu \vec\rho_\mu.\vec\tau \right) \psi 
-{1 \over 3} b m_N \sigma^3 -{1 \over 4} c \sigma^4 
+{1 \over 4}\, d\, (\omega_\alpha \omega^\alpha)
\eeq 
\vskip -0.05cm \li
The values of the parameters are collected in the following table.
\vskip -0.45cm
\begin{table}[htb]
\label{table:1}
\newcommand{\m}{\hphantom{$-$}}
\newcommand{\cc}[1]{\multicolumn{1}{c}{#1}}
\renewcommand{\tabcolsep}{1ex} 
\renewcommand{\arraystretch}{1} 
\begin{tabular}{@{}llllllll}
\hline
  & \cc{$m_\sigma$ [MeV]} & \cc{$g_\sigma$} & \cc{$g_\omega$} & \cc{$g_\rho$} &
   $b/g_\sigma^3$ & $c/g_\sigma^4$ & $d/g_\omega^4$  \\
\hline
NL3 & 508.19 & 10.217 & 12.867 & 4.4744 & $\  2.0551\, 10^{-3}$ & $-2.651\, 
10^{-3}$ & 0.0 \\
TM1 & 511.20 & 10.029 & 12.614 & 4.6322 & $-1.5082\, 10^{-3}$ 
& $\ 6.1123\, 10^{-5}$  & $2.8166\, 10^{-3}$  \\
\hline
\end{tabular}\\[2pt]
\vskip -0.2cm
\end{table}
\vskip -0.4cm
\li
These interactions do not include the $\delta$ meson considered in \cite{MP02}
nor the tensor coupling of the $\rho$ meson. The Lagrangian also does not
include a pion term; however it was shown in \cite{MP02} that the 
neutrino-nucleon scattering rate does not depend on the pion and only 
very weakly on the details of the residual contact interaction\footnote{
This statement does not hold for the emission and absorption processes
which involve charged $\rho$ and $\pi$ exchange, corrected at short distance
by contact terms.}. These parametrizations have also been shown to yield
good agreement with the available data on the longitudinal response 
for quasielastic electron scattering on nuclei in the RPA approximation.
({\it cf. e.g.} \cite{quasielastic}).
\li
We present in this contribution results for the neutrino-nucleon scattering 
rate. In this case, we need to replace the piece $D^{\rm{\small RPA}}$
describing the screened nuclear interaction by the matrix of propagators
for the neutral mesons  $\sigma$, $\omega$ and $\rho$ dressed at RPA level 
in asymmetric nuclear matter. The detailed expression for these propagators
with meson mixing in all channels was derived in \cite{M01}. 
\li
One should be aware that, in models with non linear meson couplings, 
the meson propagators are not only dressed by particle-hole loops, 
but also include a contribution of the non linearities. When performing 
the derivation of the propagators, one finds that this amounts to 
replacing the masses of the mesons appearing in the expressions of the 
propagators given in Eqs. (32-47) of \cite{M01} as follows\footnote{
The index H  indicates that the values of the meson fields are taken 
in the Hartree approximation, and $\eta^2=\eta^\mu \eta_\mu$ with 
$\eta^\mu$ defined below Eq. (6)}:
\beq
\noalign{\vskip -0.05cm}
m_\sigma^2 & \rightarrow & M_\sigma^2 = m_\sigma^2 +2\, b\, m\, \sigma_H
+3\, c\, \sigma_H^2 \nonumber \\
m_\omega^2 & \rightarrow &  M_{\omega\, T}^2 = m_\omega^2 +d\, \omega_H^2
            \quad \mbox{\rm (trans.) or}  \quad 
    M_{\omega\, L}^2 = {(m_\omega^2 
   +d\, \omega_H^2) (m_\omega^2 +3\, d\, \omega_H^2) \over m_\omega^2 
   +d\, \omega_H^2 \left( - 2\, \eta^2 +3 \right) }\quad {\rm (long.)}
 \nonumber 
\eeq
 
\section{Numerical results}
We calculated the differential and total cross sections in asymmetric
matter with a proton fraction $Y_p$. The proton fraction is determined 
by the condition that $\beta$ equilibrium is realized: $\hat\mu = 
\mu_n - \mu_p = \mu_e - \mu_\nu$. In cold neutron stars, the neutrinos 
can leave the star unhindered. On the other hand, in the conditions we 
are considering here they are still trapped inside the star, and 
the chemical potential of the neutrino therefore takes a finite value 
$\mu_\nu = (6 \pi^2 \rho Y_\nu)^{1/3}$.  The lepton fraction $Y_L=
Y_e+Y_\nu$ is determined by neutrino transport (diffusion equation 
or Boltzman), to which the cross section serves as input. A typical 
value is $Y_L \simeq 0.4$. 
\li
It is convenient to decompose the polarizations and propagators 
onto orthogonal projectors formed with the vectors and tensor available 
in the problem, {\it i.e.} the metric $g^{\mu\nu}={\rm diag}$
(1,-1,-1,-1), the hydrodynamic velocity $u^\mu$ and the transferred 
momentum $q^\mu$. 
\beq
\noalign{\vskip -0.05cm}
\Pi_{WW}^{\mu\nu\, \mbox{\rm{\small RPA}}} &=& \Pi_T\ T^{\mu\nu} 
+ \Pi_L\ \Lambda^{\mu\nu} + \Pi_Q\ Q^{\mu\nu} +i\, \Pi_E\ E^{\mu\nu} \\
\Lambda^{\mu\nu} &=& {\eta^\mu \eta^\nu \over \eta^2} \quad ; 
\quad \eta^\mu = u^\mu - {q.u \over q^2} q^\mu \qquad{\rm (longitudinal)}
\nonumber \\
T^{\mu\nu} &=& g^{\mu\nu} - {\eta^\mu \eta^\nu \over \eta^2} 
-{q^\mu q^\nu \over q^2} \qquad {\rm (transverse)} \nonumber \\
E^{\mu\nu} &=& \epsilon^{\mu\nu\rho\lambda} \eta_\rho q_\lambda 
\qquad {\rm (axial)} \quad ; \qquad
Q^{\mu\nu} = {q^\mu q^\nu \over q^2} \qquad \mbox{\rm (does not contribute)}
\nonumber
\eeq   
\vskip -0.05cm
\li The contraction of the lepton current with the polarization
can be expressed by means of three structure functions $R_1$,
$R_2$ and $R_5$ related to the previous polarizations by
\beq
\noalign{\vskip -0.05cm}
R_1 &=& {-2 \over 1-e^{-z}}{\cal I}m \left[ -{q^2 \over {\bf q^2}} 
\Pi_L + { w^2 \over  {\bf q^2}} \Pi_T \right] \quad, \qquad
R_2 = {2 \over 1-e^{-z}} {\cal I}m \left[ \Pi_T \right] \nonumber \\
R_5 &=& {2 \over 1-e^{-z}}{\cal I}m \left[ \Pi_E \right] \qquad
{\rm with} \quad z= {\omega - \Delta \mu \over k_B T}\quad
{\rm and} \quad q^\mu=(\omega,{\bf q})
\eeq
\vskip -0.05cm \li
The structure function $R_1$ involves the $\sigma$ meson 
and the longitudinal part of the $\omega$ meson, as well as the longitudinal 
part of the $\rho$. On the other hand the structure functions $R_2$ and 
$R_5$ involve the transverse part of the $\omega$ and $\rho$ mesons.
The transverse contribution is only moderately
modified by RPA correlations whereas the longitudinal contribution is 
reduced approximately by a factor of two. The contribution of the 
axial-vector response function is significant only at high density or 
energy. The relative magnitude of the contributions of the longitudinal, 
transverse and axial vector polarizations to the differential cross 
section were compared. The transverse contribution is dominant and 
will provide for about 60\% of the total result. The corrections 
will therefore arise  mostly from the subdominant longitudinal and  
axial-vector polarizations $\Pi_L$ and $\Pi_E$. 
As an exemple we show here the transverse structure 
function for the TM1 and NL3 parametrizations.
\vskip -0.3cm
\begin{figure}[htb]
\begin{minipage}[t]{80mm}
\includegraphics*[scale=0.47]{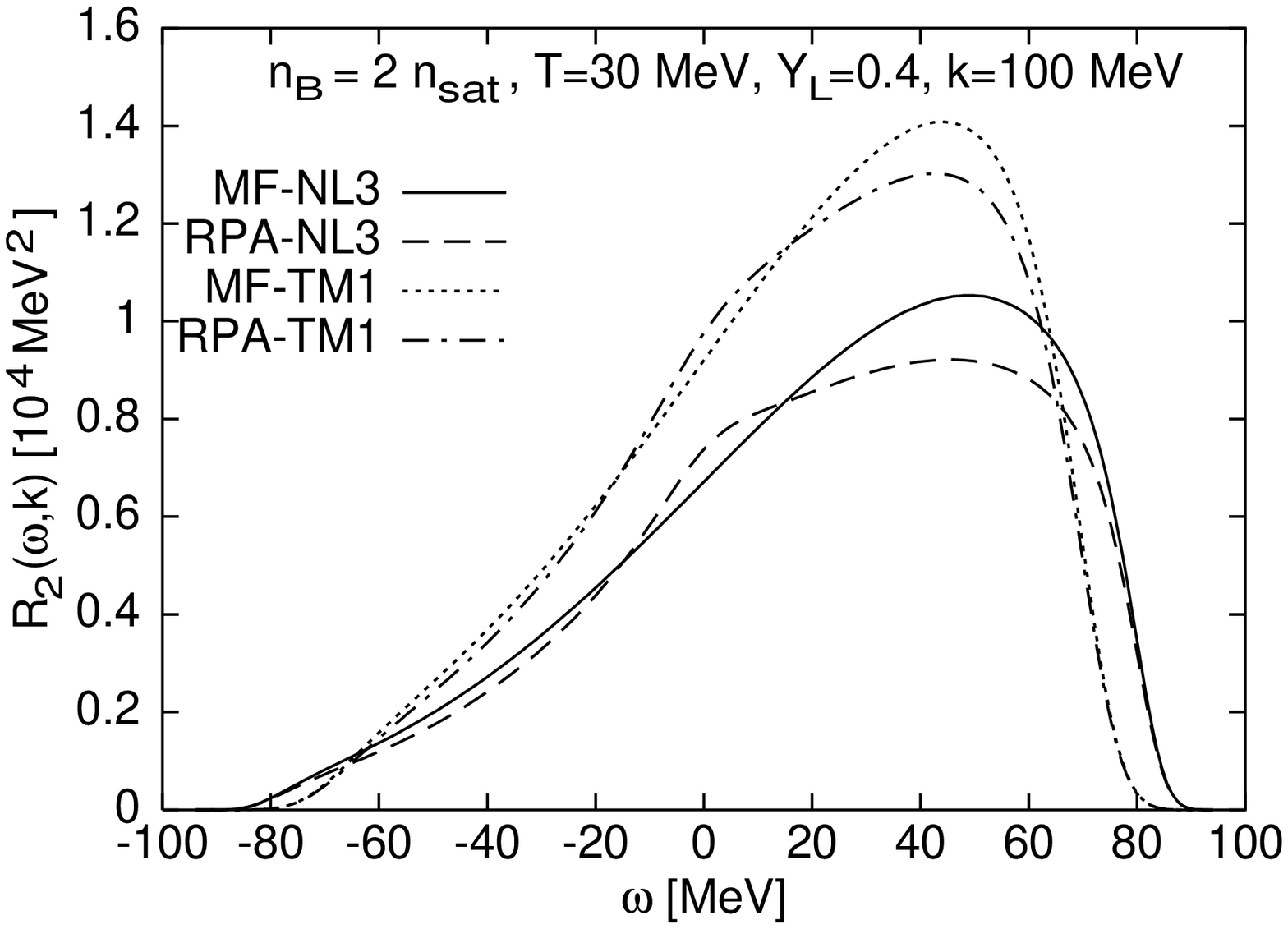}
\vskip -1cm
\caption{Transverse response function $R_2$ at twice saturation density,
temperature T=30 MeV and a fixed momentum transfer $k=100$ MeV}
\label{fig:respfun}
\end{minipage}
\hspace{\fill}
\begin{minipage}[t]{80mm}
\includegraphics*[scale=0.47]{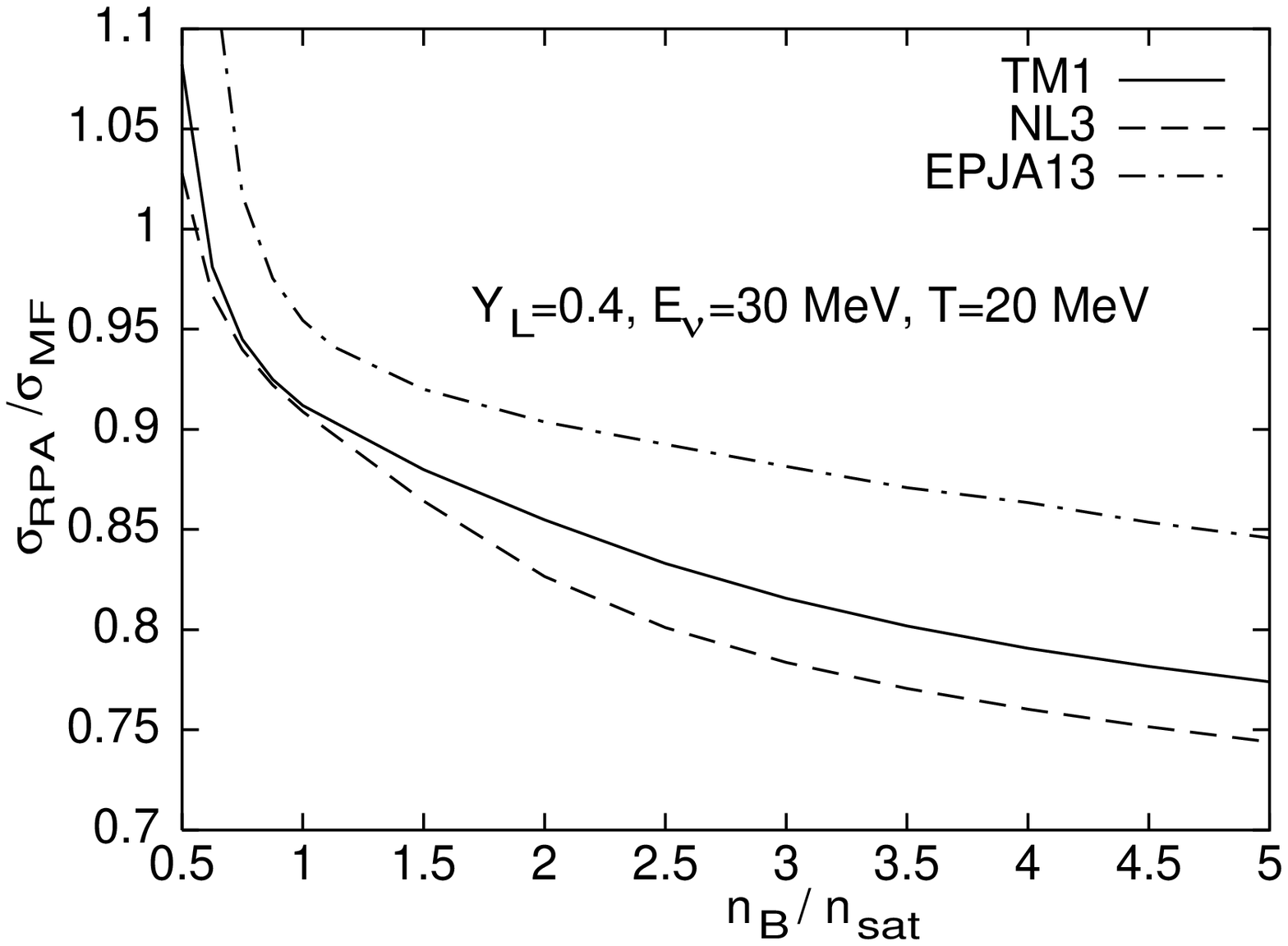}
\vskip -1cm
\caption{RPA reduction factor with respect to the mean field for the NL3
and TM1 parametrizations, compared to the reduction factor obtained in 
\cite{MP02}.}
\label{fig:reducfac}
\end{minipage}
\end{figure}
\vskip -0.3cm \li
An estimate\footnote{This is only an estimate since other processes 
contribute.} of the  mean free path is given by the inverse of the 
total cross section obtained by integrating the differential one
over the allowed energy and momentum transfer, with a Pauli blocking 
factor for the outgoing lepton. 
For the parameter sets NL3 and TM1, the total neutrino-neutron 
scattering cross section is found to be reduced by RPA correlations 
at high density by a factor $ 20 \% $ to $ 25 \% $. At low density and 
moderate temperature, on the other hand, RPA correlations would yield 
an enhancement; however the validity of the model becomes 
questionable in this range.
The reduction factor obtained here is somewhat stronger than the one
quoted in \cite{MP02}. Actually we observe that parameter sets adjusted 
to give a lower effective mass at saturation yield stronger reduction 
factors\footnote{$m^*$=0.63 for TM1 and 0.6 for NL3 while it was adjusted 
to 0.8 in \cite{MP02}}.

\end{document}